\documentclass{article}
\usepackage[T1]{fontenc} 
\usepackage[utf8]{inputenc} 
\usepackage{ismir,amsmath,cite,url}
\usepackage{graphicx}
\usepackage{color}

\usepackage{float}
\usepackage{caption}
\usepackage{subcaption}
\usepackage{tabularx}
\usepackage{multirow}
\usepackage{graphicx}
\usepackage[inline]{enumitem}
\usepackage{booktabs}
\usepackage{amssymb}
\usepackage[normalem]{ulem}
\useunder{\uline}{\ul}{}
\usepackage{tabularx}
\usepackage{lineno}

\usepackage{ulem}
\usepackage[symbol]{footmisc}

\newcommand{\correction}[1]{#1}
\newcommand{\remove}[1]{}

\title{Singer Identity Representation Learning using Self-Supervised Techniques}

\multauthor
{Bernardo Torres$^{*1}$ \hspace{1cm} Stefan Lattner$^2$ \hspace{1cm} Gaël Richard$^1$} {
 $^1$ LTCI, Telecom Paris, Institut Polytechnique de Paris\\
$^2$ Sony Computer Science Laboratories Paris\\
{\tt\small bernardo.torres@telecom-paris.fr}
}

\def\authorname{B, Torres, S. Lattner, and G. Richard}

\newcommand{\boldz}{\mathbf{z}}

\newcommand{\boldx}{\mathbf{x}}

\newcommand{\boldred}[1]{{\mathbf{#1}}}
\newcommand{\boldgreen}[1]{{\mathbf{#1}}}
\newcommand{\R}[1]{\mathbb{R}^{#1}}
\newcommand\blfootnote[1]{%
  \begingroup
  \renewcommand\thefootnote{}\footnote{#1}%
  \addtocounter{footnote}{-1}%
  \endgroup
}
\begin{document}

\maketitle

\begin{abstract}

Significant strides have been made in creating voice identity representations using speech data. However, the same level of progress has not been achieved for singing voices. To bridge this gap, we suggest a framework for training singer identity encoders to extract representations suitable for various singing-related tasks, such as singing voice similarity and synthesis. We explore different self-supervised learning techniques on a large collection of isolated vocal tracks and apply data augmentations during training to ensure that the representations are invariant to pitch and content variations. We evaluate the quality of the resulting representations on singer similarity and identification tasks across multiple datasets, with a particular emphasis on out-of-domain generalization. Our proposed framework produces high-quality embeddings that outperform both speaker verification and wav2vec 2.0 pre-trained baselines on singing voice while operating at 44.1 kHz. We release our code and trained models to facilitate further research on singing voice and related areas. 

\end{abstract}

\section{Introduction}\label{sec:introduction}

Singer representation learning is a complex task in Music Information Retrieval (MIR) that involves extracting a representation of a singer's voice, capturing their unique identity \remove{and}\correction{or} vocal timbre. This task is closely related to singer recognition, which comprises two major tasks: singer identification and singer verification. The first aims to determine the singer of a given song from a fixed set of singers in the dataset, while the latter aims to determine if two audio excerpts come from the same singer or not. Singer representation learning has many potential applications, including retrieval tasks (such as retrieving songs with a similar singing voice), and providing singer embeddings for conditioning  Singing Voice Synthesis (SVS)\cite{ wangMRSVSSingingVoice2022} and Singing Voice Conversion (SVC) systems\cite{nercessianEndtoEndZeroShotVoice2021}.
\blfootnote{*Work mostly conducted during an internship at Sony CSL Paris}

Singer recognition is related to speaker recognition, a well-established domain with vast literature. Historically, it  has received a much greater interest in particular due to the need for authentication by voice in many telecommunications applications. Singing voice, however, is different from speech in several ways, typically containing a wider variance of phoneme duration, utterances, and a wider pitch range, which makes singer recognition more challenging. Moreover, the lack of large labeled datasets further restricts the development of data-driven approaches.

In this study, we investigate if speaker recognition models trained on  \remove{speech in a supervised manner}\correction{labeled speech data} can be applied to singing voice, and whether self-supervised learning (SSL) models trained on singing voice data can achieve comparable performance. We compare different self-supervised techniques, including SimCLR \cite{chen2020simple}, Uniformity-Alignment \cite{wangUnderstandingBehaviourContrastive2021}, VICReg \cite{bardesVICRegVarianceInvarianceCovarianceRegularization2022} and BYOL \cite{grillBootstrapYourOwn2020}, trained on a large collection of isolated vocal tracks. We also explore high-frequency regions that are traditionally ignored in speech \cite{ternstromHiFiVoiceObservations2008, monsonPerceptualSignificanceHighfrequency2014} but might be present in singing voice by working in 44.1 kHz sampling rate.  Finally, we evaluate the generalization capabilities of our models on out-of-domain data. 

Our main contributions are as follows:
\begin{enumerate*}
\item We perform singer representation learning experiments using self-supervised techniques, an area that few works have explored.
\item We train encoders that operate at 44.1 kHz on a large dataset of singing voice recordings.
\item We conduct an extensive evaluation of the obtained embeddings for singer identification and singer similarity tasks, comparing them with publicly available pre-trained speech baselines.
\item We measure the out-of-domain generalization capabilities of our models on four public datasets. 
\end{enumerate*}

\section{Related work}

Singer recognition has traditionally relied on acoustic features such as Mel-frequency cepstral coefficients (MFCCs) or Line Spectral Frequencies (LSFs) to capture timbre \cite{sahidullahUsePerceptualLine2010, regnierSingerVerificationSinger2012, nakanoVocalTimbreAnalysis2014}. Some approaches focus on \correction{singer identification} on polyphonic music \cite{mesarosSingerIdentificationPolyphonic2007, lagrangeRobustSingerIdentification2012}, while others separate vocals from background \cite{sharmaImportanceAudioSourceSeparation2019, hsiehAddressingConfoundsAccompaniments2020}. In speaker verification \correction{literature}, time-invariant embeddings such as i-vector \cite{dehakFrontendFactorAnalysis2010} or x-vector \cite{snyderXVectorsRobustDNN2018} \correction{have been extensively used, and the domain} has shifted towards data-driven approaches using deep neural networks to encode acoustic features into a lower-dimensional representation that captures speaker characteristics. Temporal aggregation is used to remove the time dimension, and these systems are usually optimized \correction{using speaker label infomation} \remove{using supervised techniques like softmax}\correction{for classification} or metric learning losses. Recent works have also explored SSL for speaker verification \cite{jungPushingLimitsRaw2022, sangSelfSupervisedSpeakerVerification2022, xiaSelfsupervisedTextindependentSpeaker2021, lepageLabelefficientSelfsupervisedSpeaker2022}.

SSL has been successful in many domains, particularly with approaches such as SimCLR \cite{chen2020simple}, MoCo \cite{heMomentumContrastUnsupervised2020}, CPC \cite{oordRepresentationLearningContrastive2018}, and BYOL \cite{grillBootstrapYourOwn2020}. In the audio domain,
following the success in Computer Vision and Natural Language Processing (NLP), successful SSL models for speech include Wav2Vec 2.0 \cite{baevskiWav2vecFrameworkSelfsupervised2020}, HuBERT \cite{hsuHubertSelfsupervisedSpeech2021},  and WavLM \cite{chenWavLMLargescaleSelfsupervised2022}. SSL has also been successful in learning general-purpose audio representations, with examples like COLA \cite{saeedContrastiveLearningGeneralpurpose2021}, CLAR \cite{al-tahanClarContrastiveLearning2021}, and CLMR \cite{spijkervetContrastiveLearningMusical2021}.

 While the idea of finding singer embeddings using contrastive approaches is not new \cite{wangSingingStyleInvestigation2018}, to the best of our knowledge, only one work has employed SSL for singer representations  \cite{yakuraSelfSupervisedContrastiveLearning2022}. In their work, contrastive learning is used to acquire feature embeddings of singing voices using data augmentations that disturb a singer's identity to make the embeddings more attentive to timbre or technique. In contrast, our work explores different SSL techniques, focuses on out-of-domain testing, and evaluates on singer similarity as well as singer identification.

\section{Method} \label{sec:method}

\subsection{Goal} \label{sec:goal}

Our objective is to obtain, from isolated vocal tracks, unique singer representations  that capture the timbre of the singer's voice. These representations must satisfy three criteria:  (I) clips from the same singer should have a higher average similarity than clips from different singers; (II) the representation should not be dependent on fundamental frequency or linguistic content variations; and (III) the representations should generalize well to out-of-domain data. 

\subsection{Overview}\label{sec:overview}

The ideal embedding space for singer representations should cluster elements of the same singer while also ensuring semantic consistency by placing similar voice timbres close to each other within the space \cite{wangUnderstandingBehaviourContrastive2021}. In line with the criteria outlined in Section \ref{sec:goal}, we conducted experiments with various self-supervised techniques which force embeddings of similar input data to be close in the embedding space. We experimented with four frameworks:  SimCLR \cite{chen2020simple}, Uniformity-alignment \cite{wangUnderstandingBehaviourContrastive2021}, VICReg \cite{bardesVICRegVarianceInvarianceCovarianceRegularization2022}, and BYOL \cite{grillBootstrapYourOwn2020}. Although these frameworks share a common goal, they differ in their approach (see Section \ref{sec:loss_computation}).
 We took great care in selecting appropriate data augmentations and used a diverse set of singing voice training data.  In the current section, we describe the general training framework common to all our self-supervised experiments.


 \textbf{Data sampling}: In our methodology, we use a COLA \cite{saeedContrastiveLearningGeneralpurpose2021} approach to train our models by sampling audio segments on the fly, from a randomly drawn audio clip coming from a large database.  We first extract two segments $(x, x') \in \R{N}$ cropped randomly from the audio clip, called the anchor and positive segment. We obtain augmented views of both audio segments of the positive pair via a data augmentation module $\text{Aug}(\cdot)$ that operates in the waveform domain, resulting in an augmented positive pair $(x^{(1)}, x'^{(2)})$.  We repeat this process $B$ times for a batch size of $B$, obtaining a positive pair batch $(\boldx^{(1)}, \boldx^{(2)})$, with no repetition of audio clips during a training epoch. The superscript $'$ is further omitted for simplicity.

 \textbf{Model}:  Our proposed model takes raw audio waveforms sampled at 44.1 kHz as input. Firstly, we compute log-compressed mel-spectrogram features $m \in \R{F \times L}$ on the fly  using the \textit{nnAudio} library\footnotemark[1]. Next, the encoder module $g(\cdot)$ maps the extracted mel-spectrograms to a latent representation $h' = g(m) \in \R{H \times L}$. At this stage, adaptive average pooling is used to aggregate embedding vectors $h'$ into time-invariant feature embeddings $h \in \R{H}$. A projection layer $p(\cdot)$ maps $h$ into a lower dimensional latent space $z = p(h) \in \R{D}$ using a shallow neural network.

We denote the full model $f(\cdot)$ by stacking the acoustic feature extraction, encoder, and projection modules. During training, we encode the training batch and obtain projections $\boldz = f(\boldx$).  After training is completed, we discard the projection layer and use only the feature embeddings $h$. The similarity between a pair of embeddings is computed using the cosine similarity.

Although there are many specialized speaker verification architectures in the speech domain \cite{desplanquesECAPATDNNEmphasizedChannel2020, chungDefenceMetricLearning2020a}, we use the EfficientNet-B0 \cite{tanEfficientnetRethinkingModel2019} architecture as the backbone for the encoder module and a single SiLU non-linearity followed by a fully-connected layer for the projection layer. The projections are $\ell_2$ normalized.
\footnotetext[1]{https://github.com/KinWaiCheuk/nnAudio}

\subsection{Self-supervised frameworks} \label{sec:loss_computation}

The core concept of all used approaches is to leverage big amounts of unlabeled data to build a good representation space by aligning similar elements (and possibly separating dissimilar ones). At training time, model $f(\cdot)$ acts in a Siamese setup by encoding both elements of the augmented pair $\boldz^{(1)} = f(\boldx^{(1)}) $ and $ \boldz^{(2)} = f(\boldx^{(2)})$. For BYOL,  we have a separate encoder $f'$ with the same architecture as $f$ and we compute $\boldz^{(1)} = f(\boldx^{(1)}) $ and $ \boldz^{(2)} = f'(\boldx^{(2)})$. For all setups, we compute a loss function on the batch projections $\mathcal{L} (\boldz^{(1)}, \boldz^{(2)}$).

\textbf{Contrastive Learning}: We employ the contrastive loss called \textit{NT-Xent} from SimCLR \cite{chen2020simple}. The loss maximizes the agreement between positive samples and pushes all other embeddings of the batch (the negative parts) away in the representation space. It does so by maximizing the cosine similarity ($\text{sim}$) between positive samples and minimizing the sum of similarities for all other pairs formed in the batch:
 
\begin{equation}\label{eq:contrastive_loss_3}
\mathcal{L}_{\text{cont}}(\boldz) =
- \sum_i \log \frac{\text{exp}(\text{sim}(\boldgreen{z_i^{(1)}}, \boldgreen{z_i^{(2)}}) / \tau)}{\sum_{\boldred{j} \neq \boldgreen{i}} \text{exp}( \text{sim}(\boldgreen{z_i^{(1)}}, \boldred{z_j^{(2)}}) / \tau)}.
\end{equation}

We decouple the term $\text{exp}(\text{sim}(\boldgreen{z_i^{(1)}}, \boldgreen{z_i^{(2)}}) / \tau)$ from the denominator of the original \textit{NT-Xent} \cite{chen2020simple}, which has been shown to make the SSL task easier for smaller batch sizes and less sensitive to the hyperparameter $\tau$ \cite{yehDecoupledContrastiveLearning2022}. 

\textbf{Uniformity-Alignment}: Proposed in \cite{wangUnderstandingBehaviourContrastive2021}, Uniformity-Alignment aims to align similar examples and distribute elements uniformly in an $\ell_2$ normalized embedding space. Instead of using a contrastive loss, the authors propose optimizing directly for these two properties, resulting in two loss functions: alignment ($\mathcal{L}_{\text {align}}$) and uniformity ($\mathcal{L}_{\text {unif}}$). 

\begin{equation}\label{eq:vicreg_invariance}
\mathcal{L}_{\text {align}}(\boldz^{(1)}, \boldz^{(2)}) =\frac{1}{N} \sum_i\left\|\boldgreen{z_{i}^{(1)}}-\boldgreen{z_{i}^{(2)}}\right\|^{2},
\end{equation}

\begin{equation}\label{eq:uniformity}
    \mathcal{L}_{\text {unif }}^k(\boldgreen{z^{(k)}}) = \log \frac{1}{N} \sum_{i, j}\left(\text{exp}(-t\| \boldgreen{z_{i}^{(k)}} - \boldgreen{z_{j}^{(k)}} \|^{2})\right),
\end{equation}
where $t=2$ and $\mathcal{L}_{\text {unif }} = \sum_{k=1,2} \mathcal{L}_{\text {unif }}^k / 2 $.

\textbf{VICReg}: VICReg \cite{bardesVICRegVarianceInvarianceCovarianceRegularization2022} is an approach that attempts to maximize the information content of the learned embeddings. Three losses are proposed: the variance, invariance, and covariance losses. The invariance loss is the same as the alignment loss (see Equation \ref{eq:vicreg_invariance}). The variance regularization forces the standard deviation of a batch (in the dimension axis) to be close to the value $\mu$, preventing \textit{collapse} (when embedding dimensions become useless). Let $\mathbf{d_j(z)} \in \mathbb{R}^B$ be the vector composed of the values of a batch $\mathbf{z}$ at dimension j. The variance regularization is:

\begin{equation}\label{eq:vicreg_variance}
    \mathcal{L}_{\text {var}} (\boldz)=\frac{1}{D} \sum_{j=1}^{D} \max \left(0, \mu -S\left(\mathbf{d_j(z)}, \epsilon\right)\right),
\end{equation}
where D is the number of dimensions of $z_i$, and $S$ is the regularized standard deviation  $S(x, \epsilon)=\sqrt{\operatorname{Var}(x)+\epsilon}$.

The covariance regularization decorrelates the dimensions of the embedding, making them orthogonal:

\begin{equation}\label{eq:vicreg_covariance}
    \mathcal{L}_{\text {cov}}(\boldz)=\frac{1}{D_z} \sum_{\boldgreen{i} \neq \boldred{j}}(C(\boldz))_{\boldgreen{i}, \boldred{j}}^{2},
\end{equation}
where  $C(\boldz)=\frac{1}{N-1} \sum_{i=1}^{N}\left(\boldz_{i}-\bar{\boldz}\right)\left(\boldz_{i}-\bar{\boldz}\right)^{T}$ is the covariance matrix of $ \boldz$, and $\bar{\boldz}=\frac{1}{N} \sum_{i=1}^{N} \boldz_{i}$.

\textbf{BYOL}: Bootstrap Your Own Latent (BYOL) \cite{grillBootstrapYourOwn2020} employs two neural networks: the online and target networks. Both networks share the same architecture. In addition, BYOL employs an additional predictor network $q$ which computes predictions $q(\boldz)$.   BYOL iteratively refines the representation of the online network by minimizing the mean squared error (MSE) between its predictions and the target's projections. If $f$ and $f'$ denote the online and target networks, respectively, the loss function $\mathcal{L}_{\text {BYOL}}$ on the projections $\boldz^{(1)} = f(\boldx^{(1)})$, $\boldz^{(2)} = f'(\boldx^{(2)})$ is:

\begin{equation}\label{eq:byol}
\mathcal{L}_{\text {BYOL}}(\boldz^{(1)}, \boldz^{(2)}) =\frac{1}{N} \sum_i\left\|\boldgreen{z_{i}^{(1)}}-q(\boldgreen{z_{i}^{(2)}})\right\|^{2}.
\end{equation}

The target network $f'$ is not trained using directly the gradients of $\mathcal{L}_{\text {BYOL}}$, but it is updated with  an exponential moving average of the weights of the online network.

\section{Experiments}

\subsection{Data} \label{sec:data}

\textbf{} We used a large \correction{private} corpus of professionally recorded singing voice data containing approximately 25,000 tracks, \remove{from 5,700 artists} totaling 940 hours of audio data. The dataset \correction{consists of isolated vocals of re-recordings of popular songs by 5,700 artists} and includes a variety of singing styles, voice types, lyrics, and audio effects. \correction{We note that the actual number of singers is unknown, as the same artist might have been re-recorded by multiple singers. Therefore, we do not believe that this corpus is appropriate for supervised training}. Additionally, we added 6 hours of source-separated vocals to the corpus. All samples were converted to mono 44.1kHz tracks with 16-bit encoding, and any silence lasting more than 1.3 seconds was trimmed to 1.3 seconds. Segments with less than 0.5\% amplitude were considered silent, and segments with more than 0.5\% amplitude lasting less than 0.2 seconds were silenced. The dataset was partitioned into three distinct sets with ratios of 80\% for training, 10\% for validation, and 10\% for testing, with no \remove{singer}\correction{artist} allocated to more than one set. The length of a track is typically a few minutes.

\textbf{Out-of-domain evaluation}:
Four datasets are used to test the out-of-domain generalization of the models. The summary of all datasets is shown in Table \ref{tab:ood_datasets}.

\begin{table}[]
\centering
\resizebox{\columnwidth}{!}{%
\begin{tabular}{lllll}
\toprule
\textbf{Corpus} & \textbf{Language} & \textbf{\#Hours} & \textbf{\#Singers} & \textbf{Type}  \\ \midrule
VCTK \cite{yamagishiCSTRVCTKCorpus2019}             & English           & 44               & 110                & Speech         \\
NUS-48E \cite{duanNUSSungSpoken2013}        & English           & 1.91             & 12                 & Speech/Singing \\
VocalSet \cite{wilkinsVocalSetSingingVoice2018}        & English           & 10.1             & 20                 & Singing        \\
M4Singer \cite{zhangM4SingerMultistyleMultisinger2022}        & Chinese           & 29.77            & 20                 & Singing        \\ \bottomrule
\end{tabular}%
}
\caption{Out-of-domain datasets used for testing.}
\label{tab:ood_datasets}
\end{table}

\subsection{Experiment setup}\label{sec:exp_setup}

We perform a series of experiments to determine the best SSL framework for singer representation learning: 
\begin{itemize}

    \item \texttt{CONT}: We train a model on the decoupled version of the contrastive loss  $\mathcal{L} = \mathcal{L}_{\text {cont}}$ \cite{yehDecoupledContrastiveLearning2022}.
    \item \texttt{CONT-VC}: We train a model using $\mathcal{L}_{\text {cont}}$ (contrastive loss) with added variance and covariance regularization $\mathcal{L} =  \mathcal{L}_{\text {cont}} + \mu \mathcal{L}_{\text {var}} + \nu \mathcal{L}_{\text {cov}} $ \cite{lattnerSampleMatchDrumSample2022}.
    
    \item \texttt{UNIF}: We train a model using the uniformity- alignment loss $\mathcal{L} = \mathcal{L}_{\text {align}} + \gamma \mathcal{L}_{\text {unif}}$ \cite{wangUnderstandingBehaviourContrastive2021}.
    \item \texttt{VICReg}: We train a model using the VICReg loss $\mathcal{L} = \lambda \mathcal{L}_{\text {align}} + \mu \mathcal{L}_{\text {var}} + \nu \mathcal{L}_{\text {cov}} $ \cite{bardesVICRegVarianceInvarianceCovarianceRegularization2022} .
    \item \texttt{BYOL}: We train a model on BYOL configuration, optimizing the MSE $\mathcal{L} = \mathcal{L}_{\text {BYOL}}$  \cite{grillBootstrapYourOwn2020}. 

\end{itemize}

The contrastive loss has been shown to yield good results in the literature \cite{saeedContrastiveLearningGeneralpurpose2021, yakuraSelfSupervisedContrastiveLearning2022}, but there is \remove{also} concern that it may break the semantic structure of the embeddings by pushing similar singers away in the representation space \cite{wangUnderstandingBehaviourContrastive2021}. In the \texttt{CONT-VC} approach, the addition of variance and covariance losses from VICReg is tested as a regularization method to mitigate this problem \cite{lepageLabelefficientSelfsupervisedSpeaker2022, lattnerSampleMatchDrumSample2022}. The \texttt{UNIF} approach attempts to optimize directly for uniformity of the space, which has shown links with linear separability \cite{wangUnderstandingBehaviourContrastive2021} and potential for strong singer identification results. While \texttt{VICReg} claims to be an information-theoretic approach to general-purpose representation learning, it has not yet been thoroughly tested in the audio domain. Finally, \texttt{BYOL} is included in the study for comparison as it has shown promising results in several audio downstream tasks, claiming state-of-the-art\cite{niizumiBYOLAudioExploring2023}.




\subsection{Evaluation procedure} \label{sec:eval_procedure}

The models are first trained until the validation loss stops decreasing. Validation similarity metrics (Section \ref{sec:singer_similarity}) are tracked during training, and the best-performing model is selected. This model is evaluated on the test and on out-of-domain sets using cropped 4-second clips of singer recordings (with no overlapping segments). \correction{The embeddings $h$ are evaluated in two tasks: singer/speech similarity and singer/speech identification. For simplicity, we refer to singer similarity/identification even when dealing with speech data such as with VCTK/NUS-48E datasets (see Section \ref{sec:discussion}} for details).
\subsubsection{Singer similarity} \label{sec:singer_similarity}

We evaluate singer similarity by measuring two metrics 
\remove{In this work, we evaluate singer similarity through two metrics, measured} directly on the singer feature embeddings $h$: the Equal Error Rate (EER) and  Mean Normalized Rank (MNR). The EER relates to singer verification. On the other hand, we relate the MNR to singer retrieval by computing the similarities between a query excerpt and a set of candidates. No training is performed for the similarity evaluation.

\textbf{EER}: The EER is a popular metric for verification systems\remove{that measures the error rate for a binary classifier}. To compute the EER, the system is exposed to a set of trials consisting of true pairs (two segments coming from the same singer) and fake pairs (two segments coming from different singers), and a similarity metric is computed for both cases (in our case the cosine similarity). False positives (FP) and False Negatives (FN) can be computed by applying a threshold $\tau$ on the similarity metric, and the Detection Error Tradeoff (DET) is obtained by varying $\tau$ as a function of FP and FN. The EER is the error rate at which FP = FN. We compute the EER following the implementation available as part of the SupERB benchmark \cite{yangSUPERBSpeechProcessing2021}\footnotemark[2]. We sample 50,000 speaker pairs for computing the EER on the test set and 20,000 speaker pairs for out-of-domain.

\footnotetext[2]{https://github.com/s3prl/s3p}


\textbf{MNR}: Denote $q^{(1)}$, $q^{(2)}$ two query audio samples, coming from the same audio recording (and therefore the same singer) drawn at random at each trial. Let $S$ be a set of $N$ audio samples, drawn at random from a dataset, and $q^{(2)} \in S$. The MNR is \cite{lattnerSampleMatchDrumSample2022}:
\begin{equation}
    \text{MNR}  = \frac{1}{K} \sum_{k=1}^K \frac{R(q^{(1)}_k, S_k)}{N},
\end{equation}
where $R(q^{(1)}_k, S_k)$ is the integer position (rank) of $q^{(2)}$ in the sorted list of cosine similarities between $q^{(1)}$ and the samples in $S$. We perform  $K=1000$ trials for $N=512$.

\textbf{Input sample rate}: To ensure a fair comparison with the baselines, which operate on 16 kHz, the evaluation is done in two scenarios: at 16 kHz and 44.1 kHz. In the former, the 44.1 kHz inputs are downsampled to 16 kHz and upsampled back to 44.1 kHz before being fed to the models, removing energy above 8 kHz. In the latter, the trained models have access to the full frequency range of the input data.
\subsubsection{Singer identification}

To evaluate the linear separability of singer classes, we perform singer classification as a downstream task for singer identification on out-of-domain evaluations. We use 5-fold cross-validation to split the audio files of each singer into train, validation, and test subsets (4-fold for NUS-48E). A single feed-forward linear layer is trained with cross-entropy loss on the train subset to predict singer classes from embeddings extracted from frozen models. The best model is selected on the validation subset. Average metrics on the test set over all folds are reported. We limit this task to out-of-domain evaluations since these datasets contain multiple files per singer and the classes are balanced.

 \footnotetext[3]{https://github.com/resemble-ai/Resemblyzer}
\footnotetext[4]{https://github.com/clovaai/voxceleb\_trainer}
\footnotetext[5]{https://huggingface.co/facebook/wav2vec2-base}
\footnotetext[6]{https://huggingface.co/facebook/wav2vec2-large-xlsr-53}

\begin{table}[]
\centering
\resizebox{\columnwidth}{!}{%
\begin{tabular}{lllll}
\toprule
\begin{tabular}[c]{@{}l@{}}Model\end{tabular} & \textbf{\#Params} & \textbf{SR} & \textbf{Dim.} & \textbf{Backbone} \\ \midrule
\texttt{GE2E} \cite{wanGeneralizedEndtoendLoss2018}\footnotemark[3]        &  1.4M &  16  & 256     & LSTM                      \\
\texttt{F-ResNet} \cite{chungDefenceMetricLearning2020a}\footnotemark[4]    & 1.4M & 16 & 512     & ResNet-34  \\
\texttt{H/ASP} \cite{kwonInsOutsSpeaker2021}\footnotemark[4]  & 8.0M & 16  & 512     & ResNet-34         \\
\texttt{Wav2Vec-base} \cite{baevskiWav2vecFrameworkSelfsupervised2020}\footnotemark[5]  & 95M & 16  & 12X768 & Wav2Vec 2.0    \\
\texttt{XLSR-53} \cite{conneauUnsupervisedCrosslingualRepresentation2021}\footnotemark[6]       & 300M & 16 & 24X1024 & Wav2Vec 2.0 
\\ \midrule
Ours       & 5.0M & 44.1 & 1000 & EfficientNet-B0               \\
\bottomrule
\end{tabular}%
}
\caption{Number of network parameters,  sampling rate in kHz (SR), the size of the feature embeddings (Dim), and the architecture backbone for the baselines and our models.}
\label{tab:baselines}
\end{table}

\begin{table*}
\centering
\setlength{\tabcolsep}{3.8pt}
\resizebox{\textwidth}{!}{%
\begin{tabular}{llclclclclclclclclclc}
\toprule
& 
\multicolumn{4}{c}{\textbf{In-domain}} &
\multicolumn{16}{c}{\textbf{Out-of-domain}}
\\
\cmidrule(lr){2-5}
\cmidrule(lr){6-21}
\multicolumn{1}{c}{\textbf{}} & \multicolumn{4}{c}{\textbf{Test dataset}$^{\ast}$} & \multicolumn{4}{c}{\textbf{VCTK}} & \multicolumn{4}{c}{\textbf{NUS-48E}} & \multicolumn{4}{c}{\textbf{M4Singer}$^{\ast}$} & \multicolumn{4}{c}{\textbf{Vocalset}$^{\ast}$} \\ 
\cmidrule(lr){2-5}
\cmidrule(lr){6-9}
\cmidrule(lr){10-13}
\cmidrule(lr){14-17}
\cmidrule(lr){18-21}
\textbf{Model} & \multicolumn{2}{c}{\textbf{EER}} & \multicolumn{2}{c}{\textbf{MNR}} & \multicolumn{2}{c}{\textbf{EER}} & \multicolumn{2}{c}{\textbf{MNR}} & \multicolumn{2}{c}{\textbf{EER}} & \multicolumn{2}{c}{\textbf{MNR}} & \multicolumn{2}{c}{\textbf{EER}} & \multicolumn{2}{c}{\textbf{MNR}} & \multicolumn{2}{c}{\textbf{EER}} & \multicolumn{2}{c}{\textbf{MNR}} \\
\cmidrule(lr){2-3}
\cmidrule(lr){4-5}
\cmidrule(lr){6-7}
\cmidrule(lr){8-9}
\cmidrule(lr){10-11}
\cmidrule(lr){12-13}
\cmidrule(lr){14-15}
\cmidrule(lr){16-17}
\cmidrule(lr){18-19}
\cmidrule(lr){20-21}

 & \multicolumn{1}{c}{\textbf{16}} & \textbf{44.1} & \multicolumn{1}{c}{\textbf{16}} & \textbf{44.1} & \multicolumn{1}{c}{\textbf{16}} & \textbf{44.1} & \multicolumn{1}{c}{\textbf{16}} & \textbf{44.1} & \multicolumn{1}{c}{\textbf{16}} & \textbf{44.1} & \multicolumn{1}{c}{\textbf{16}} & \textbf{44.1} & \multicolumn{1}{c}{\textbf{16}} & \textbf{44.1} & \multicolumn{1}{c}{\textbf{16}} & \textbf{44.1} & \multicolumn{1}{c}{\textbf{16}} & \textbf{44.1} & \multicolumn{1}{c}{\textbf{16}} & \textbf{44.1} \\ \midrule
\texttt{GE2E}$^{\dagger}$ &
  27.24 &
  - &
  18.9 &
  - &
  13.42 &
  - &
  5.41 &
  - &
  28.04 &
  - &
  18.99 &
  - &
  25.01 &
  - &
  15.99 &
  - &
  40.45 &
  - &
  35.34 &
  - \\
\texttt{F-ResNet}$^{\dagger}$ &
  15.21 &
  \textbf{-} &
  7.76 &
  - &
  1.01 &
  - &
  \textbf{\underline{0.08}}$^{\star}$ &
  - &
  15.36 &
  - &
  6.63 &
  - &
  14.21 &
  - &
  5.98 &
  - &
  40.64 &
  - &
  33.82 &
  - \\
\texttt{H/ASP}$^{\dagger}$ &
  12.36 &
  - &
  5.82 &
  - &
  \textbf{\underline{0.28}}$^{\star}$ &
  - &
  \textbf{\underline{0.08}}$^{\star}$ &
  - &
  \textbf{\underline{13.99}} &
  - &
  \textbf{\underline{5.42}} &
  - &
  \underline{12.31} &
  - &
  \underline{3.93} &
  - &
  36.27 &
  - &
  30.79 &
  - \\
   \midrule
\texttt{Wav2Vec-base} &
  25.36 &
  - &
  14.78 &
  - &
  23.15 &
  - &
  15.78 &
  - &
  32.65 &
  - &
  24.39 &
  - &
  26.28 &
  - &
  13.37 &
  - &
  39.34 &
  - &
  34.23 &
  - \\
\texttt{XLSR-53} &
  25.22 &
  - &
  15.82 &
  - &
  25.93 &
  - &
  19.95 &
  - &
  36.62 &
  - &
  28.52 &
  - &
  26.02 &
  - &
  16.96 &
  - &
  40.09 &
  - &
  35.32 &
  - \\ \midrule
\texttt{VICReg} &
  8.19 &
  3.88 &
  2.29 &
  1.14 &
  25.17 &
  \multicolumn{1}{l}{23.88} &
  14.99 &
  \multicolumn{1}{l}{14.62} &
  26.11 &
  \multicolumn{1}{l}{26.06} &
  15.43 &
  \multicolumn{1}{l}{15.34} &
  24.6 &
  \multicolumn{1}{l}{22.05} &
  9.78 &
  \multicolumn{1}{l}{8.69} &
  34.58 &
  \multicolumn{1}{l}{33.12} &
  28.21 &
  \multicolumn{1}{l}{26.5} \\
\texttt{UNIF} &
  9.48 &
  2.86 &
  2.13 &
  0.78 &
  22.51 &
  \multicolumn{1}{l}{24.28} &
  12.99 &
  \multicolumn{1}{l}{14.67} &
  27.65 &
  \multicolumn{1}{l}{26.12} &
  17.08 &
  \multicolumn{1}{l}{15.48} &
  20.46 &
  \multicolumn{1}{l}{17.03} &
  8.83 &
  \multicolumn{1}{l}{6.67} &
  32.4 &
  \multicolumn{1}{l}{31.19} &
  25.07 &
  \multicolumn{1}{l}{23.19} \\
\texttt{CONT} &
  6.39 &
  \textbf{2.16} &
 \underline{1.33} &
  \textbf{0.48} &
  20.04 &
  \multicolumn{1}{l}{22.87} &
  9.34 &
  \multicolumn{1}{l}{11.56} &
  23.67 &
  \multicolumn{1}{l}{24.51} &
  12.86 &
  \multicolumn{1}{l}{12.45} &
  14.28 &
  \multicolumn{1}{l}{12.67} &
  5.52 &
  \multicolumn{1}{l}{4.51} &
  32.16 &
  \multicolumn{1}{l}{30.61} &
  23.64 &
  \multicolumn{1}{l}{22.6} \\
\texttt{CONT-VC} &
  7.39 &
  2.74 &
  1.61 &
  0.52 &
  19.92 &
  \multicolumn{1}{l}{21.79} &
  10.35 &
  \multicolumn{1}{l}{11.12} &
  24.99 &
  \multicolumn{1}{l}{25.4} &
  15.06 &
  \multicolumn{1}{l}{13.91} &
  15.97 &
  \multicolumn{1}{l}{12.68} &
  6.94 &
  \multicolumn{1}{l}{4.81} &
  \underline{31.03} &
  \multicolumn{1}{l}{\textbf{29.74}} &
  \underline{22.65} &
  \multicolumn{1}{l}{21.87} \\
\texttt{BYOL} &
  \underline{5.88} &
  3.82 &
  1.5 &
  0.68 &
  17.44 &
  \multicolumn{1}{l}{19.97} &
  7.8 &
  \multicolumn{1}{l}{9.73} &
  26.01 &
  \multicolumn{1}{l}{23.9} &
  15.62 &
  \multicolumn{1}{l}{12.21} &
  15.65 &
  \multicolumn{1}{l}{\textbf{12.28}} &
  5.86 &
  \multicolumn{1}{l}{\textbf{3.77}} &
  31.59 &
  \multicolumn{1}{l}{29.76} &
  23.93 &
  \multicolumn{1}{l}{\textbf{21.25}} \\
  \bottomrule
\end{tabular}%
}
\caption{EER and MNR (\%, lower is better) measured on frozen model embeddings. Datasets that contain only singing voice are marked with $^{\ast}$, and models which are not self-supervised are indicated with $^{\dagger}$. Results in bold are the best among all models, for both 44.1 kHz and 16 kHz input sample rates. Underlined results highlight the best on 16kHz input only. For \texttt{Wav2Vec-base} and \texttt{XLSR-53}, we use the embeddings of the first layer and aggregate them using average pooling.}\label{tab:similarity_results}
\end{table*}

\subsection{Baselines}

    In our experiments, \remove{we use several pre-trained models as baselines for singer verification tasks. We use}we use as baselines three speaker verification networks: \texttt{GE2E} \cite{wanGeneralizedEndtoendLoss2018}, Fast-ResNet34 \cite{chungDefenceMetricLearning2020a} (hereafter referred to as \texttt{F-ResNet}), \texttt{H/ASP} \cite{kwonInsOutsSpeaker2021}; and two large \correction{general purpose} self-supervised models   
    \texttt{Wav2Vec-base} \cite{baevskiWav2vecFrameworkSelfsupervised2020}, and \texttt{XLSR-53} \cite{conneauUnsupervisedCrosslingualRepresentation2021}. These models have been pre-trained on speech and either achieved state-of-the-art results or \remove{are common}\correction{have been} used for obtaining speaker representations for speech/singing voice synthesis tasks while being publicly available. We provide an overview of the baseline models in Table \ref{tab:baselines}.
    
    Since all baselines\remove{are pre-trained on speech data and} operate on 16kHz, we downsample the test signals to 16kHz accordingly. For \texttt{Wav2Vec-base} and \texttt{XLSR-53}, we use adaptive average pooling as the temporal aggregation method for the \correction{frame-wise} feature embeddings, and we employ a learned, weighted sum of the first three layers for the downstream classifier \cite{yangSUPERBSpeechProcessing2021}. We empirically found that this approach boosts classification performance compared to using a single layer. Specifically, the first layers of these models are more effective for speaker verification \cite{chenWavLMLargescaleSelfsupervised2022} and are more correlated with speaker characteristics \cite{choiNeuralAnalysisSynthesis2021}. For singer similarity evaluations, we use only the first layer, as there is no training involved to yield weights for a weighted sum. 



\subsection{Training} \label{sec:training}

To train our models, we used $4$-second audio clips that were normalized, augmented, and converted to log-compressed mel-filterbanks with $80$ mel bins, a window length of $2048$, and a hop size of $512$. This results in an FFT frame of $46.4$ms and sliding windows of $11.6$ms for 44.1 kHz audio. We initialized the EfficientNet-B0 backbone with pre-trained weights on ImageNet \cite{lattnerSampleMatchDrumSample2022} and used the ADAM optimizer with a learning rate of 1e-4 and weight decay of 1e-5, with a batch size of 120. For contrastive loss, we used a temperature parameter of $\tau=0.2$ \cite{wangUnderstandingBehaviourContrastive2021}, and whenever we used covariance regularization, we set $\nu=100$. For variance regularization, we set $\mu=25$. Additionally, for \texttt{VICReg} experiments, we used an invariance loss factor of $\lambda=25$, and \texttt{UNIF}, we set $\gamma=1$. For \texttt{BYOL}, we used a learning rate of 3e-5, a weight decay of 1.5e-6 and an initial moving average value $\tau$ of 0.99. We found through empirical analysis that these hyperparameters were effective for convergence and avoiding collapse.

In terms of data augmentation techniques, we applied Gaussian noise, gain with a minimum attenuation of -6 dB, and time masking with at most 1/8 of the clip being masked. We also used formant-preserving pitch shifting with Praat \cite{boersmaPraatDoingPhonetics2009, jadoulIntroducingParselmouthPython2018} as a method of data augmentation, with the pitch shift ratio and pitch range ratio being sampled uniformly from U(1,3) and U(1,1.5), respectively, with a random choice on whether to take the reciprocal of the sampled ratios or not \cite{choiNeuralAnalysisSynthesis2021}. All augmentations had a probability of 0.5 of being applied. We avoided using naive pitch-shifting techniques that transpose the formants, which can significantly alter the singers' timbre.

\begin{table}
\centering
\resizebox{\columnwidth}{!}{%
\begin{tabular}{lllll}
\toprule
\textbf{Model} & \textbf{VCTK} & \multicolumn{1}{c}{\textbf{NUS-48E}\phantom{}} & \multicolumn{1}{c}{\textbf{M4Singer}$^{\ast}$} & \multicolumn{1}{c}{\textbf{Vocalset}$^{\ast}$} \\ \midrule
\texttt{GE2E}$^{\dagger}$        & 97.01 & 91.13          & 88.72 & 45.66          \\
\texttt{F-ResNet}$^{\dagger}$   & 99.91          & 97.36          & 94.51 & 49.52          \\
\texttt{H/ASP}$^{\dagger}$ & \textbf{99.93} & \textbf{98.32} & 97.87 & 74.65          \\
\midrule
\texttt{Wav2Vec-base}  &     98.70         &  96.16         &   96.52    & 79.19          \\
\texttt{XLSR-53}  &        99.66         &      97.02          &   \textbf{98.62}    &     \textbf{86.05}\\ \midrule
\texttt{VICReg}       & 52.52          & 78.98          & 87.34 & 49.69          \\
\texttt{UNIF}         & 74.43          & 93.05          & 93.55 & 67.52          \\
\texttt{CONT}         & 90.24          & 96.23          & 95.72 & 77.42          \\
\texttt{CONT-VC}   & 86.03          & 95.14          & 94.69 & 75.20          \\
\texttt{BYOL}         &     \underline{96.95}           & \underline{96.56} &    \underline{97.00}   & \underline{81.01} \\  \bottomrule
\end{tabular}%
}
\caption{Average linear classification accuracy on out-of-domain data (\%) over K-fold cross-validation. Datasets that contain only singing voice are marked with $\ast$. The best scores are highlighted in bold and the best among the trained models (bottom 5 rows) are underlined. Models which are not self-supervised are indicated with $\dagger$.}
\label{tab:classification_results}
\end{table}

\section{Results and Discussion} \label{sec:discussion}

Table \ref{tab:similarity_results} presents the results of singer similarity evaluation on both in-domain and out-of-domain test sets, reporting the best Equal Error Rate (EER) and Mean Normalized Rank (MNR) for trained models and baselines in all test datasets.  Table \ref{tab:classification_results} shows the accuracies for downstream singer identification task on out-of-domain datasets. We also share in supplementary material additional qualitative visual evaluations of the embeddings\footnotemark[7], and release code and models to encourage reproducibility and facilitate its use in future projects\footnotemark[8].

\footnotetext[7]{https://sites.google.com/view/singer-representation-learning}
\footnotetext[8]{https://github.com/SonyCSLParis/ssl-singer-identity}

\subsection{Results on pre-trained models on speech}

The results indicate that models pre-trained on speech in a supervised manner (using speaker labels)  exhibit good generalization to out-of-domain speech datasets. \texttt{H/ASP} achieves an impressive 0.28\% EER on the VCTK, and all models score higher than 88\% accuracy on VCTK, NUS-48E, and M4Singer datasets. Their similarity performance on singing voice datasets, however, is much worse than on speech, but the best \remove{supervised} models still score below 10\% EER on NUS-48E and 12.31\% and 14.21\% EER on M4Singer for \texttt{H/ASP} and \texttt{F-ResNet}, respectively.

This suggests that important features of the singing voice can also be learned directly from speech. However, the results show the pre-trained models perform worse on heavily processed data that includes uncommon effects and vocal techniques. This is evident, in particular, in the last columns of Tables \ref{tab:similarity_results} and \ref{tab:classification_results} (VocalSet), with all baselines scoring around 40\% EER and from 20\% to 40\% worse accuracy when compared to the other datasets.

\subsection{Results on Self-Supervised Models}

 Models trained with contrastive loss (\texttt{CONT} and \texttt{CONT-VC)} achieved the best EER and MNR on the test set.  These models were able to learn highly discriminative features for the task of in-domain singer similarity. For instance, the \texttt{CONT} model had the lowest overall EER and  MNR (2.16\% and 0.48\% respectively) on the test set.

It can also be seen in Table \ref{tab:similarity_results} that in-domain test performance did not necessarily translate to good generalization to out-of-domain data. By adding variance and covariance regularizations (\texttt{CONT-VC}), the model achieved better generalization to out-of-domain data on some datasets (such as the VocalSet, with approximately 1\% EER difference). However, in the \texttt{VICReg} scenario, which has both regularizations but lacks the contrastive part, the results were worse. In fact, \texttt{VICReg} had the worst overall results of all the tested self-supervised frameworks. \correction{\texttt{UNIF}, while better than \texttt{VICReg}, also performed worse on average when compared to the other approaches. }

\texttt{CONT} and \texttt{BYOL} achieved the best accuracy over all our trained models on singer identification (Table \ref{tab:classification_results}), achieving the highest scores of 77.42\% and 81.01\%, respectively (the VocalSet paper \cite{wilkinsVocalSetSingingVoice2018} reports 60-70\% accuracy on a supervised singer identification task).

\texttt{BYOL} achieved the best generalization on similarity, performing best on out-of-domain data, even though its scores were worse on the in-domain test set. Interestingly, of all explored self-supervised techniques, \texttt{BYOL} is the only one that does not explicitly force any kind of feature distribution on the embedding space. In addition, \texttt{BYOL} was able to learn best how to leverage the information present at 16 kHz sample rate, with an EER of 5.88 on the test set. It also performed best on out-of-domain speech data (VCTK). In general, our models struggled with speech, performing generally better when they only had access to a reduced frequency band. This suggests that in speech, high-frequency information the models rely on hinders their ability to generalize.

    \textbf{44.1 vs 16 kHz}: Using 44.1 kHz inputs consistently improved the similarity results on singing voice datasets (e.g., M4Singer) for all models, \correction{highlighting the models' ability to efficiently use high-frequency information}. Moreover, most models showed a marked \correction{decline in the in-domain dataset results when tested with 16 kHz inputs (the \texttt{CONT} model, for example, shows a drop from 2.16\% EER to 6.39\% EER). While the 16 kHz inputs could be considered out-of-domain, this effect shows that high-frequency information is important for the trained models to achieve better performance.}

    \textbf{Comparison to baselines}: The trained models show better results than baselines on in-domain test sets and the VocalSet dataset for singer similarity tests, although they fall behind \texttt{F-ResNet} and \texttt{H/ASP} on the mixed speech/singing dataset NUS-48E and VCTK. Nonetheless, on M4Singer, some self-supervised models outperformed the supervised baselines, with \texttt{BYOL} showing the best performance (12.28\% EER and 3.77\% MNR)\correction{, and \texttt{CONT} and \texttt{CONT-VC} also being superior to \texttt{F-ResNet}. }

The trained models have substantially better singer similarity results compared to \texttt{Wav2Vec-base} and \texttt{XLSR-53}. These results indicate the potential of training models on the proposed SSL tasks specifically on singing voice data. Further improvements could be made by fine-tuning the embeddings on verification tasks, as has been demonstrated in previous work on Wav2Vec 2.0 \cite{fanExploringWav2vecSpeaker2021}.

Moreover, \texttt{BYOL} outperformed \texttt{Wav2Vec-base} for both VocalSet and M4Singer on classification. Among all models, \texttt{XLSR-53} achieved the best overall performance for singer identification of singing voice. However, is noteworthy that our models have significantly fewer parameters than the self-supervised \texttt{Wav2Vec-base} (19 times less) and \texttt{XLSR-53} (63 times less).

\section{Conclusion}




In conclusion, we have shown that self-supervised learning is an effective approach for learning representations of singers. The self-supervised models trained on a large corpus of singing voice data demonstrated a performance that either matched or surpassed publicly available supervised speech models, without resorting to specialized architecture designs. Additionally, our models outperformed general-purpose self-supervised counterparts even with a significantly reduced parameter count. When applied to singer identification, our models exhibited superior performance over \texttt{Wav2vec-base} on singing voice datasets but fell somewhat short in comparison to the considerably more expansive \texttt{XLSR-53}.

Furthermore, our results suggest that these models hold promise for singer identification and similarity downstream tasks. \texttt{BYOL} showed the most promise for generalizing to out-of-domain data, while the contrastive approaches were more effective for in-domain data.

However, we note that our models' representations do not yet fully capture a singer's identity when confronted with unique singing techniques, such as those found in the VocalSet \cite{wilkinsVocalSetSingingVoice2018}. This underscores the need for further research on robust SSL frameworks capable of accommodating such variations. Our findings also suggest that employing a higher sampling frequency can be advantageous for singing voice tasks, but optimal frequency for generalizing to both singing and speech tasks remains to be determined.

\section{Acknowledgments}
This work was partly funded by the European Union (ERC, HI-Audio, 101052978). Views and opinions expressed are however those of the author(s) only and do not necessarily reflect those of the European Union or the European Research Council. Neither the European Union nor the granting authority can be held responsible for them.

\correction{The authors also thank Alain Riou for the BYOL implementation.}

\bibliography{ISMIRtemplate} 

\begin{thebibliography}{10}
\providecommand{\url}[1]{#1}
\csname url@samestyle\endcsname
\providecommand{\newblock}{\relax}
\providecommand{\bibinfo}[2]{#2}
\providecommand{\BIBentrySTDinterwordspacing}{\spaceskip=0pt\relax}
\providecommand{\BIBentryALTinterwordstretchfactor}{4}
\providecommand{\BIBentryALTinterwordspacing}{\spaceskip=\fontdimen2\font plus
\BIBentryALTinterwordstretchfactor\fontdimen3\font minus
  \fontdimen4\font\relax}
\providecommand{\BIBforeignlanguage}[2]{{%
\expandafter\ifx\csname l@#1\endcsname\relax
\typeout{** WARNING: IEEEtran.bst: No hyphenation pattern has been}%
\typeout{** loaded for the language `#1'. Using the pattern for}%
\typeout{** the default language instead.}%
\else
\language=\csname l@#1\endcsname
\fi
#2}}
\providecommand{\BIBdecl}{\relax}
\BIBdecl

\bibitem{wangMRSVSSingingVoice2022}
S.~Wang, J.~Liu, Y.~Ren, Z.~Wang, C.~Xu, and Z.~Zhao, ``{{MR-SVS}}: {{Singing}}
  voice synthesis with multi-reference encoder,'' \emph{CoRR}, vol.
  abs/2201.03864, 2022.

\bibitem{nercessianEndtoEndZeroShotVoice2021}
S.~Nercessian, ``End-to-{{End Zero-Shot Voice Conversion Using}} a {{DDSP
  Vocoder}},'' in \emph{{{IEEE Workshop}} on {{Applications}} of {{Signal
  Processing}} to {{Audio}} and {{Acoustics}}}, Oct. 2021, pp. 1--5.

\bibitem{chen2020simple}
T.~Chen, S.~Kornblith, M.~Norouzi, and G.~Hinton, ``A simple framework for
  contrastive learning of visual representations,'' in \emph{International
  Conference on Machine Learning}, 2020, pp. 1597--1607.

\bibitem{wangUnderstandingBehaviourContrastive2021}
F.~Wang and H.~Liu, ``Understanding the {{Behaviour}} of {{Contrastive
  Loss}},'' in \emph{2021 {{IEEE}}/{{CVF Conference}} on {{Computer Vision}}
  and {{Pattern Recognition}} ({{CVPR}})}.\hskip 1em plus 0.5em minus
  0.4em\relax {Nashville, TN, USA}: {IEEE}, Jun. 2021, pp. 2495--2504.

\bibitem{bardesVICRegVarianceInvarianceCovarianceRegularization2022}
A.~Bardes, J.~Ponce, and Y.~LeCun, ``{{VICReg}}:
  {{Variance-invariance-covariance}} regularization for self-supervised
  learning,'' in \emph{{{ICLR}}}, 2022.

\bibitem{grillBootstrapYourOwn2020}
J.-B. Grill, F.~Strub, F.~Altch{\'e}, C.~Tallec, P.~H. Richemond,
  E.~Buchatskaya, C.~Doersch, B.~{\'A}. Pires, Z.~Guo, M.~G. Azar, B.~Piot,
  K.~Kavukcuoglu, R.~Munos, and M.~Valko, ``Bootstrap your own latent - {{A}}
  new approach to self-supervised learning,'' in \emph{{{NeurIPS}}}, 2020.

\bibitem{ternstromHiFiVoiceObservations2008}
S.~Ternstr{\"o}m, ``Hi-{{Fi}} voice: Observations on the distribution of energy
  in the singing voice spectrum above 5 {{kHz}},'' in \emph{Acoustics' 08,
  Paris, France, Jun 29-{{Jul}} 4, 2008}, 2008, pp. 3171--3176.

\bibitem{monsonPerceptualSignificanceHighfrequency2014}
B.~B. Monson, E.~J. Hunter, A.~J. Lotto, and B.~H. Story, ``The perceptual
  significance of high-frequency energy in the human voice,'' \emph{Frontiers
  in psychology}, vol.~5, p. 587, 2014.

\bibitem{sahidullahUsePerceptualLine2010}
{\relax Md}.~Sahidullah, S.~Chakroborty, and G.~Saha, ``On the use of
  perceptual {{Line Spectral}} pairs {{Frequencies}} and higher-order residual
  moments for {{Speaker Identification}},'' \emph{IJBM}, vol.~2, no.~4, p. 358,
  2010.

\bibitem{regnierSingerVerificationSinger2012}
L.~Regnier and G.~Peeters, ``Singer verification: {{Singer}} model .vs. song
  model,'' in \emph{{{IEEE International Conference}} on {{Acoustics}},
  {{Speech}} and {{Signal Processing}} ({{ICASSP}})}.\hskip 1em plus 0.5em
  minus 0.4em\relax {Kyoto, Japan}: {IEEE}, 2012, pp. 437--440.

\bibitem{nakanoVocalTimbreAnalysis2014}
T.~Nakano, K.~Yoshii, and M.~Goto, ``Vocal timbre analysis using latent
  {{Dirichlet}} allocation and cross-gender vocal timbre similarity,'' in
  \emph{{{IEEE International Conference}} on {{Acoustics}}, {{Speech}} and
  {{Signal Processing}} ({{ICASSP}})}.\hskip 1em plus 0.5em minus 0.4em\relax
  {IEEE}, 2014, pp. 5202--5206.

\bibitem{mesarosSingerIdentificationPolyphonic2007}
A.~Mesaros, T.~Virtanen, and A.~Klapuri, ``Singer identification in polyphonic
  music using vocal separation and pattern recognition methods.'' in
  \emph{Proc. of the 8th International Society for Music Information Retrieval
  Conference ({{ISMIR}})}, 2007, pp. 375--378.

\bibitem{lagrangeRobustSingerIdentification2012}
M.~Lagrange, A.~Ozerov, and E.~Vincent, ``Robust singer identification in
  polyphonic music using melody enhancement and uncertainty-based learning,''
  in \emph{Proc. of the 13th International Society for Music Information
  Retrieval Conference ({{ISMIR}})}, 2012.

\bibitem{sharmaImportanceAudioSourceSeparation2019}
B.~Sharma, R.~K. Das, and H.~Li, ``On the {{Importance}} of {{Audio-Source
  Separation}} for {{Singer Identification}} in {{Polyphonic Music}},'' in
  \emph{Interspeech 2019}.\hskip 1em plus 0.5em minus 0.4em\relax {ISCA}, Sep.
  2019, pp. 2020--2024.

\bibitem{hsiehAddressingConfoundsAccompaniments2020}
T.-H. Hsieh, K.-H. Cheng, Z.-C. Fan, Y.-C. Yang, and Y.-H. Yang, ``Addressing
  the confounds of accompaniments in singer identification,'' in \emph{{{IEEE
  International Conference}} on {{Acoustics}}, {{Speech}} and {{Signal
  Processing}} ({{ICASSP}})}, 2020, pp. 1--5.

\bibitem{dehakFrontendFactorAnalysis2010}
N.~Dehak, P.~J. Kenny, R.~Dehak, P.~Dumouchel, and P.~Ouellet, ``Front-end
  factor analysis for speaker verification,'' \emph{IEEE Transactions on Audio,
  Speech, and Language Processing}, vol.~19, no.~4, pp. 788--798, 2010.

\bibitem{snyderXVectorsRobustDNN2018}
D.~Snyder, D.~{Garcia-Romero}, G.~Sell, D.~Povey, and S.~Khudanpur,
  ``X-{{Vectors}}: {{Robust DNN Embeddings}} for {{Speaker Recognition}},'' in
  \emph{{{IEEE International Conference}} on {{Acoustics}}, {{Speech}} and
  {{Signal Processing}} ({{ICASSP}})}.\hskip 1em plus 0.5em minus 0.4em\relax
  {IEEE}, Apr. 2018, pp. 5329--5333.

\bibitem{jungPushingLimitsRaw2022}
J.-w. Jung, Y.~J. Kim, H.-S. Heo, B.-J. Lee, Y.~Kwon, and J.~S. Chung,
  ``Pushing the limits of raw waveform speaker recognition,'' in
  \emph{Interspeech}.\hskip 1em plus 0.5em minus 0.4em\relax {ISCA}, 2022, pp.
  2228--2232.

\bibitem{sangSelfSupervisedSpeakerVerification2022}
M.~Sang, H.~Li, F.~Liu, A.~O. Arnold, and L.~Wan, ``Self-supervised speaker
  verification with simple siamese network and self-supervised
  regularization,'' in \emph{{{IEEE International Conference}} on
  {{Acoustics}}, {{Speech}} and {{Signal Processing}} ({{ICASSP}})}.\hskip 1em
  plus 0.5em minus 0.4em\relax {IEEE}, 2022, pp. 6127--6131.

\bibitem{xiaSelfsupervisedTextindependentSpeaker2021}
W.~Xia, C.~Zhang, C.~Weng, M.~Yu, and D.~Yu, ``Self-supervised text-independent
  speaker verification using prototypical momentum contrastive learning,'' in
  \emph{{{IEEE International Conference}} on {{Acoustics}}, {{Speech}} and
  {{Signal Processing}} ({{ICASSP}})}.\hskip 1em plus 0.5em minus 0.4em\relax
  {IEEE}, 2021, pp. 6723--6727.

\bibitem{lepageLabelefficientSelfsupervisedSpeaker2022}
T.~Lepage and R.~Dehak, ``Label-efficient self-supervised speaker verification
  with information maximization and contrastive learning,'' in \emph{Proc.
  {{Interspeech}} 2022}.\hskip 1em plus 0.5em minus 0.4em\relax {ISCA}, Sep.
  2022, pp. 4018--4022.

\bibitem{heMomentumContrastUnsupervised2020}
K.~He, H.~Fan, Y.~Wu, S.~Xie, and R.~Girshick, ``Momentum contrast for
  unsupervised visual representation learning,'' in \emph{Proc. of the
  {{IEEE}}/{{CVF}} Conference on Computer Vision and Pattern Recognition},
  2020, pp. 9729--9738.

\bibitem{oordRepresentationLearningContrastive2018}
A.~van~den Oord, Y.~Li, and O.~Vinyals, ``Representation learning with
  contrastive predictive coding,'' \emph{arXiv preprint arXiv:1807.03748},
  2018.

\bibitem{baevskiWav2vecFrameworkSelfsupervised2020}
A.~Baevski, Y.~Zhou, A.~Mohamed, and M.~Auli, ``Wav2vec 2.0: {{A}} framework
  for self-supervised learning of speech representations,'' \emph{Advances in
  neural information processing systems}, vol.~33, pp. 12\,449--12\,460, 2020.

\bibitem{hsuHubertSelfsupervisedSpeech2021}
W.-N. Hsu, B.~Bolte, Y.-H.~H. Tsai, K.~Lakhotia, R.~Salakhutdinov, and
  A.~Mohamed, ``Hubert: {{Self-supervised}} speech representation learning by
  masked prediction of hidden units,'' \emph{IEEE/ACM Transactions on Audio,
  Speech, and Language Processing}, vol.~29, pp. 3451--3460, 2021.

\bibitem{chenWavLMLargescaleSelfsupervised2022}
S.~Chen, C.~Wang, Z.~Chen, Y.~Wu, S.~Liu, Z.~Chen, J.~Li, N.~Kanda,
  T.~Yoshioka, X.~Xiao \emph{et~al.}, ``Wavlm: {{Large-scale}} self-supervised
  pre-training for full stack speech processing,'' \emph{IEEE Journal of
  Selected Topics in Signal Processing}, vol.~16, no.~6, pp. 1505--1518, 2022.

\bibitem{saeedContrastiveLearningGeneralpurpose2021}
A.~Saeed, D.~Grangier, and N.~Zeghidour, ``Contrastive learning of
  general-purpose audio representations,'' in \emph{{{IEEE International
  Conference}} on {{Acoustics}}, {{Speech}} and {{Signal Processing}}
  ({{ICASSP}})}.\hskip 1em plus 0.5em minus 0.4em\relax {IEEE}, 2021, pp.
  3875--3879.

\bibitem{al-tahanClarContrastiveLearning2021}
H.~{Al-Tahan} and Y.~Mohsenzadeh, ``Clar: {{Contrastive}} learning of auditory
  representations,'' in \emph{International Conference on Artificial
  Intelligence and Statistics}, 2021, pp. 2530--2538.

\bibitem{spijkervetContrastiveLearningMusical2021}
J.~Spijkervet and J.~A. Burgoyne, ``Contrastive learning of musical
  representations,'' in \emph{Proc. of the 22nd International Society for Music
  Information Retrieval Conference ({{ISMIR}})}, 2021, pp. 673--681.

\bibitem{wangSingingStyleInvestigation2018}
C.-i. Wang and G.~Tzanetakis, ``Singing {{Style Investigation}} by {{Residual
  Siamese Convolutional Neural Networks}},'' in \emph{{{IEEE International
  Conference}} on {{Acoustics}}, {{Speech}} and {{Signal Processing}}
  ({{ICASSP}})}, Apr. 2018, pp. 116--120.

\bibitem{yakuraSelfSupervisedContrastiveLearning2022}
H.~Yakura, K.~Watanabe, and M.~Goto, ``Self-{{Supervised Contrastive Learning}}
  for {{Singing Voices}},'' \emph{IEEE/ACM Trans. Audio Speech Lang. Process.},
  vol.~30, pp. 1614--1623, 2022.

\bibitem{desplanquesECAPATDNNEmphasizedChannel2020}
B.~Desplanques, J.~Thienpondt, and K.~Demuynck, ``{{ECAPA-TDNN}}:
  {{Emphasized}} channel attention, propagation and aggregation in {{TDNN}}
  based speaker verification,'' in \emph{Interspeech}.\hskip 1em plus 0.5em
  minus 0.4em\relax {ISCA}, 2020, pp. 3830--3834.

\bibitem{chungDefenceMetricLearning2020a}
J.~S. Chung, J.~Huh, S.~Mun, M.~Lee, H.~S. Heo, S.~Choe, C.~Ham, S.~Jung, B.-J.
  Lee, and I.~Han, ``In defence of metric learning for speaker recognition,''
  in \emph{Interspeech 2020}, Oct. 2020, pp. 2977--2981.

\bibitem{tanEfficientnetRethinkingModel2019}
M.~Tan and Q.~Le, ``Efficientnet: {{Rethinking}} model scaling for
  convolutional neural networks,'' in \emph{International Conference on Machine
  Learning}, 2019, pp. 6105--6114.

\bibitem{yehDecoupledContrastiveLearning2022}
C.-H. Yeh, C.-Y. Hong, Y.-C. Hsu, T.-L. Liu, Y.~Chen, and Y.~LeCun, ``Decoupled
  contrastive learning,'' in \emph{{{ECCV}} (26)}, ser. Lecture Notes in
  Computer Science, vol. 13686.\hskip 1em plus 0.5em minus 0.4em\relax
  {Springer}, 2022, pp. 668--684.

\bibitem{yamagishiCSTRVCTKCorpus2019}
J.~Yamagishi, C.~Veaux, and K.~MacDonald, ``{{CSTR VCTK Corpus}}: {{English
  Multi-speaker Corpus}} for {{CSTR Voice Cloning Toolkit}} (version 0.92),''
  2019.

\bibitem{duanNUSSungSpoken2013}
Z.~Duan, H.~Fang, B.~Li, K.~C. Sim, and Y.~Wang, ``The {{NUS}} sung and spoken
  lyrics corpus: {{A}} quantitative comparison of singing and speech,'' in
  \emph{{{APSIPA}}}.\hskip 1em plus 0.5em minus 0.4em\relax {IEEE}, 2013, pp.
  1--9.

\bibitem{wilkinsVocalSetSingingVoice2018}
J.~Wilkins, P.~Seetharaman, A.~Wahl, and B.~Pardo, ``{{VocalSet}}: {{A Singing
  Voice Dataset}}.'' in \emph{Proc. of the 19th International Society for Music
  Information Retrieval Conference ({{ISMIR}})}, 2018, pp. 468--474.

\bibitem{zhangM4SingerMultistyleMultisinger2022}
L.~Zhang, R.~Li, S.~Wang, L.~Deng, J.~Liu, Y.~Ren, J.~He, R.~Huang, J.~Zhu,
  X.~Chen \emph{et~al.}, ``{{M4Singer}}: {{A}} multi-style, multi-singer and
  musical score provided mandarin singing corpus,'' \emph{Advances in Neural
  Information Processing Systems}, vol.~35, pp. 6914--6926, 2022.

\bibitem{lattnerSampleMatchDrumSample2022}
S.~Lattner, ``{{SampleMatch}}: {{Drum}} sample retrieval by musical context,''
  in \emph{Proc. of the 23rd International Society for Music Information
  Retrieval Conference ({{ISMIR}})}, 2022, pp. 781--788.

\bibitem{niizumiBYOLAudioExploring2023}
D.~Niizumi, D.~Takeuchi, Y.~Ohishi, N.~Harada, and K.~Kashino, ``{{BYOL}} for
  audio: {{Exploring}} pre-trained general-purpose audio representations,''
  \emph{IEEE ACM Trans. Audio Speech Lang. Process.}, vol.~31, pp. 137--151,
  2023.

\bibitem{yangSUPERBSpeechProcessing2021}
S.-W. Yang, P.-H. Chi, Y.-S. Chuang, C.-I.~J. Lai, K.~Lakhotia, Y.~Y. Lin,
  A.~T. Liu, J.~Shi, X.~Chang, G.-T. Lin, T.-H. Huang, W.-C. Tseng, K.-t. Lee,
  D.-R. Liu, Z.~Huang, S.~Dong, S.-W. Li, S.~Watanabe, A.~Mohamed, and H.-y.
  Lee, ``{{SUPERB}}: {{Speech}} processing universal {{PERformance}}
  benchmark,'' in \emph{Interspeech}.\hskip 1em plus 0.5em minus 0.4em\relax
  {ISCA}, 2021, pp. 1194--1198.

\bibitem{wanGeneralizedEndtoendLoss2018}
L.~Wan, Q.~Wang, A.~Papir, and I.~L. Moreno, ``Generalized end-to-end loss for
  speaker verification,'' in \emph{{{IEEE International Conference}} on
  {{Acoustics}}, {{Speech}} and {{Signal Processing}} ({{ICASSP}})}.\hskip 1em
  plus 0.5em minus 0.4em\relax {IEEE}, 2018, pp. 4879--4883.

\bibitem{kwonInsOutsSpeaker2021}
Y.~Kwon, H.-S. Heo, B.-J. Lee, and J.~S. Chung, ``The ins and outs of speaker
  recognition: Lessons from {{VoxSRC}} 2020,'' in \emph{{{IEEE International
  Conference}} on {{Acoustics}}, {{Speech}} and {{Signal Processing}}
  ({{ICASSP}})}.\hskip 1em plus 0.5em minus 0.4em\relax {IEEE}, 2021, pp.
  5809--5813.

\bibitem{conneauUnsupervisedCrosslingualRepresentation2021}
A.~Conneau, A.~Baevski, R.~Collobert, A.~Mohamed, and M.~Auli, ``Unsupervised
  cross-lingual representation learning for speech recognition,'' in
  \emph{Interspeech}.\hskip 1em plus 0.5em minus 0.4em\relax {ISCA}, 2021, pp.
  2426--2430.

\bibitem{choiNeuralAnalysisSynthesis2021}
H.-S. Choi, J.~Lee, W.~Kim, J.~Lee, H.~Heo, and K.~Lee, ``Neural analysis and
  synthesis: {{Reconstructing}} speech from self-supervised representations,''
  \emph{Advances in Neural Information Processing Systems}, vol.~34, pp.
  16\,251--16\,265, 2021.

\bibitem{boersmaPraatDoingPhonetics2009}
P.~Boersma and D.~Weenink, ``Praat: Doing phonetics by computer ({{Version}}
  5.1.13),'' 2009.

\bibitem{jadoulIntroducingParselmouthPython2018}
Y.~Jadoul, B.~Thompson, and B.~{de Boer}, ``Introducing {{Parselmouth}}: {{A
  Python}} interface to {{Praat}},'' \emph{Journal of Phonetics}, vol.~71, pp.
  1--15, Nov. 2018.

\bibitem{fanExploringWav2vecSpeaker2021}
Z.~Fan, M.~Li, S.~Zhou, and B.~Xu, ``Exploring wav2vec 2.0 on {{Speaker
  Verification}} and {{Language Identification}},'' in \emph{Interspeech
  2021}.\hskip 1em plus 0.5em minus 0.4em\relax {ISCA}, Aug. 2021, pp.
  1509--1513.

\end{thebibliography}

%
%
%
%
%

\end{document}